\documentclass[aps,onecolumn,prd,showpacs,nofootinbib]{revtex4}
\usepackage{amsmath}
\usepackage{graphicx}
\usepackage{subfigure}
\usepackage{dcolumn}
\usepackage{bm}
\usepackage{amssymb}
\usepackage{latexsym}

\newcommand{\be}{\begin{equation}}
\newcommand{\en}{\end{equation}}
\newcommand{\bea}{\begin{eqnarray}}
\newcommand{\ena}{\end{eqnarray}}

\newcommand{\mP}{m_{_{\mathrm{Pl}}}}

\newcommand{\mean}[1]{\left\langle #1 \right\rangle}
\newcommand{\mpl}{m_{_\mathrm{Pl}}}

\newcommand{\ie}{\textsl{i.e.}~}

\begin{document}

\title{Moduli Fields as Quintessence and
the Chameleon}

\author{Philippe Brax} \email{brax@spht.saclay.cea.fr}
\affiliation{ Service de Physique Th\'eorique,
CEA-Saclay, Gif/Yvette cedex, France F-91191}

\author{J\'er\^ome Martin} \email{jmartin@iap.fr} \affiliation{
Institut d'Astrophysique de Paris, UMR 7095-CNRS, Universit\'e Pierre
et Marie Curie, 98bis boulevard Arago, 75014 Paris, France}

\date{\today}

\begin{abstract}
We consider models where moduli fields are not stabilized and play the
role of quintessence. In order to evade gravitational tests, we
investigate the possibility that moduli behave as chameleon fields. We
find that, for realistic moduli superpotentials, the chameleon effect
is not strong enough, implying that moduli quintessence models are
gravitationally ruled out. More generally, we state a no-go theorem
for quintessence in supergravity whereby models either behave like a
pure cosmological constant or violate gravitational tests.
\end{abstract}

\pacs{98.80.Cq, 98.70.Vc}
\maketitle

\section{Introduction}

Dark energy and its properties is one of the most intriguing puzzles
of present day theoretical physics. Indeed, there is convincing
evidence, coming from SNIa supernovae~\cite{IA}, large scale
structures of the universe~\cite{LSS, SMB, FA} and the CMB
anisotropies~\cite{CMB, SMB2} which leads to the existence of an
acceleration of the universe expansion in the recent past. When
interpreted within the realm of General Relativity, these results
imply the existence of a pervading weakly interacting fluid with a
negative equation of state and a dominant energy density. The simplest
possibility is of course a pure cosmological constant. This has the
advantage of both fitting the data and incorporating a mild
modification of the Einstein equations. Now it happens that the value
of the cosmological constant is so small compared to high energy
physics scales that no proper explanation for such a fine--tuning has
been found except maybe the anthropic principle~\cite{Wein} used in
the context of a stringy landscape~\cite{Susskind, Pol}. This is all
the more puzzling in view of the very diverse sources of radiative
corrections in the standard model of particle physics and beyond.

\par

A plausible alternative involves the presence of a scalar field akin
to the inflaton of early universe cosmology and responsible for the
tiny vacuum energy scale~\cite{RP,
quint,PB,BM1,BM2,BMR1,BMR2,MSU}. These models of quintessence have
nice features such as the presence of long time attractors (tracking
fields) leading to a relative insensitivity to initial
conditions~\cite{RP}.  In most cases, the quintessence runaway
potentials lead to large values of the quintessence field now, of the
order of the Planck mass. This immediately prompts the necessity of
embedding such models in high energy physics where nearly Planck scale
physics is taken into account. The most natural possibility is
supergravity as it involves both supersymmetry and gravitational
effects~\cite{Nilles}. Moreover, superstring theories lead to
supergravity models at low energy.

\par

From the model building point of view, the quintessence field does not
belong to the well-known sector of particles of the standard
model. Therefore, one has to envisage a dark sector where this field
lives and provide the corresponding K\"ahler, $K_{\rm quint}$, and
super potentials $W_{\rm quint}$ in order to compute the quintessence
scalar potential explicitly. Once a quintessence model has been built,
one must also worry about the coupling to both matter and hidden
sector supersymmetry breaking~\cite{BMpart}. Indeed the rolling of the
quintessence field can induce variations of constants such as the fine
structure constants. Moreover the smallness of the mass of the
quintessence field implies that its gravitational coupling to matter
must be suppressed in order to comply with fifth force and equivalence
principle violation experiments~\cite{GR,DP}.

\par

The observable sector is fairly well-known and the hidden sector can
be parameterized. Therefore, the main uncertainty comes from the dark
sector, \ie from the specific form chosen for $K_{\rm quint}$ and
$W_{\rm quint}$. Recently, we have investigated this question for a
class of models where the K\"ahler potential and the superpotential
can be Taylor expanded or are given by polynomial functions of the
(super) fields~\cite{BMcosmo}. We have shown that this type of models,
under the standard assumption of separate sectors (see also our
conclusion), is in trouble as either they are uninteresting from the
cosmological point of view (typically, in practice, they are
equivalent to a cosmological constant) or they violate the bounds from
gravity experiments (typically, they violate the bound on the fifth
force and/or on the weak equivalence principle).

\par

The aim of this paper is to study a general class of models, probably
the most natural one from a string theory point of
view~\cite{noscaleST}, where the quintessence field is a moduli field
(K\"ahler moduli). Technically, this means that $K_{\rm quint}$ is
taken to be a logarithm of the quintessence
field~\cite{noscaleST}. Although the K\"ahler function is known, there
is no specific standard choice for the superpotential which remains a
free function. Therefore, we will derive model independent results and
then discuss the various cases that have been envisaged in the
literature for $W_{\rm quint}$ (for instance, polynomial
superpotentials and exponential ones {\em \`a la}
KKLT~\cite{KKLT}). We show that, for reasonable choices of $W_{\rm
quint}$, the corresponding models are also in trouble from the gravity
experiments point of view. This last result is in fact more subtle
than in the case of the first class of models treated in
Ref.~\cite{BMcosmo}. Indeed, contrary to the polynomial models, a
chameleon mechanism~\cite{cham} can be present in the no scale case
and could be used to protect the quintessence field from gravity
problems. However, unfortunately, we show that this mechanism is in
fact not sufficiently efficient to save no scale quintessence in
simple cases such as gaugino condensation and polynomial
superpotentials.

\par

The paper is arranged as follows. In Sec.~\ref{sec:noscale}, we
establish some general results relevant to the no-scale models. In
particular, in sub-Sec.~\ref{subsec:scalar}, we calculate the
quintessence potential for a general moduli superpotential and in
sub-Sec.~\ref{subsec:soft}, we give the corresponding soft terms in
the observable sector. In sub-Sec.~\ref{subsec:electroweak}, we study
how the electroweak transition is affected by the no scale dark
sector. Then, in Sec.~\ref{sec:testcham}, we briefly review the
chameleon mechanism. In particular, in sub-Sec.~\ref{subsec:cham}, we
describe the thin shell phenomenon with, in
sub-Sec.~\ref{subsec:kkltpot}, applications to the gaugino
condensation case and in sub-Sec.~\ref{subsec:nrpot} to the polynomial
case. In Sec.~\ref{sec:conclusion}, we present our conclusions and
state a no-go theorem for the compatibility between quintessence in
supergravity and gravity experiments.

\section{No Scale Quintessence}
\label{sec:noscale}

\subsection{The Scalar Potential}
\label{subsec:scalar}

In this section we collect results related to the dynamics of K\"ahler
moduli coming from string compactifications. In practice we only
consider that there is a single moduli $Q$ which can be seen as the
breathing mode of the compactification manifold. The reduction from
$10$ dimensions to $4$ dimensions leads to a no-scale structure for
the K\"ahler potential of the moduli. The K\"ahler potential is given
by the following expression
\begin{equation}
K_{\rm quint} =-\frac{3}{\kappa }\ln \left[\kappa
  ^{1/2}\left(Q+Q^{\dagger }\right)\right]\, ,
\end{equation}
where $\kappa \equiv 8\pi /\mpl^2$. The moduli $Q$ has no potential
and is a flat direction to all order in perturbation theory. In string
theory, the validity of the supergravity approximation is guaranteed
provided $\kappa^{1/2} Q \gg 1$, implying that the compactification
manifold is larger than the string scale. A potential can be generated
once non-perturbative effects are taken into account, this may lead to
a superpotential
\begin{equation}
W_{\rm quint}=W_{\rm quint}(Q)\equiv M^3{\cal W}\left(\kappa
^{1/2}Q\right)\, .
\end{equation}
which will be discussed later. The advantage of the above writing is
that it emphasizes the scale $M$ of the superpotential. The quantity
${\cal W}$ is dimensionless and of order one. Then, inserting the
K\"ahler and the super potentials into the expression of the scalar
potential, one gets
\begin{equation}
\label{potquint}
V_{\rm quint}(Q)=-\frac{\kappa ^{1/2}}{\left[\kappa
^{1/2}\left(Q+Q^{\dagger }\right)\right]^2}\left(W\frac{{\partial
}W^{\dagger }}{\partial Q^{\dagger }}+W^{\dagger }\frac{{\partial
}W}{\partial Q}\right) +\frac{1}{3\kappa ^{1/2}\left(Q+Q^{\dagger
}\right)}\left\vert \frac{{\partial }W}{\partial Q}\right\vert ^2\, .
\end{equation}
The noscale property implies that the term in $-3 \vert W\vert ^2$ in
the supergravity potential cancels. The kinetic terms of the moduli
read $3\vert \partial Q\vert ^2/\left(Q+Q^{\dagger}\right)^2$ implying
that $Q$ is not a normalized field. The normalized field $q$ is given
by
\begin{equation}
\kappa^{1/2} Q=\exp\left(-\sqrt{\frac{2}{3}} q\right)\, .
\end{equation}
where $q$ is a dimensionless scalar field.

\par

As soon as a quintessence field has a runaway potential and leads to
the present day acceleration of the universe expansion, its mass is
tiny and may lead to gravitational problems. In order to minimize this
problem, we assume that the quintessence sector is only coupled
gravitationally to the observable and hidden sectors~\cite{BMpart}. In
some sense, this assumption is that of non triviality of the
model. The corresponding situation can be described by the following
K\"ahler and super potentials~\cite{BMpart}
\begin{equation}
K= K_{\rm quint} + K_{\rm hid} + K_{\rm obs},\ W= W_{\rm quint} +
W_{\rm hid}+  W_{\rm obs}\, .
\end{equation}
Now the observable sector is known since it comprises the fields of
the Minimal Standard Supersymetric Model (MSSM) $\phi^a$ and the
corresponding superpotential can be expressed as~\cite{Nilles}
\begin{equation}
W_{\rm obs}= \frac{1}{2}  \mu_{ab} \phi^a \phi^b +\frac{1}{3}
\lambda_{abc} \phi^a \phi^b \phi^c \, ,
\end{equation}
where $\mu_{ab}$ is a supersymmetric mass matrix and
$\lambda_{abc}$ the Yukawa couplings.

\par

The fact that susy is broken in an hidden sector modifies the shape of
the quintessence potential. Another way to put it is that the susy
breaking causes the appearance of soft terms in the dark sector and
these soft terms are responsible for the modification of the
quintessence potential. The new shape has been computed in
Ref.~\cite{BMpart}. If we parametrise the hidden sector supersymmetry
breaking in a model independent way, we have
\begin{equation}
\label{parahidden} \kappa ^{1/2}\mean{z_i}_{\rm min}\sim a_i(Q)\,
, \quad \kappa
    \mean{W_{\rm hid}}_{\rm min}\sim M_{_{\rm S}}(Q)\, , \quad \kappa
    ^{1/2}\mean{\frac{\partial W_{\rm hid}}{\partial z_i}}_{\rm
    min}\sim c_i(Q)M_{_{\rm S}}(Q)\, ,
\end{equation}
where $a_i$ and $c_i$ are coefficients whose values depend on the
detailed structure of the hidden sector. Notice that the coupling of
the hidden sector to the quintessence sector implies that the vev's of
the hidden sector fields responsible for supersymmetry breaking can
depend on the quintessence field. Taking into account the no scale
shape of the K\"ahler potential, one finds
\begin{eqnarray}
\label{potDE}
V_{_{\rm DE}} &=& {\rm e}^{\sum _i\vert a_i\vert ^2}\kappa M^6 \left\{
\frac{1}{\left[\kappa ^{1/2}\left(Q+Q^{\dagger
}\right)\right]^2}\left[{\cal W}\frac{{\partial }{\cal W}^{\dagger
}}{\partial \left(\kappa ^{1/2}Q^{\dagger }\right)}+{\cal W}^{\dagger
}\frac{{\partial }{\cal W}}{\partial \left(\kappa
^{1/2}Q\right)}\right] +\frac{1}{3\kappa ^{1/2}\left(Q+Q^{\dagger
}\right)}\left\vert \frac{{\partial }{\cal W}}{\partial \left(\kappa
^{1/2}Q\right)}\right\vert ^2\right\} \nonumber \\
& & -M_{_{\rm S}}M^3
\frac{{\rm e}^{\sum _i\vert a_i\vert ^2}}{\left[\kappa ^{1/2}\left(Q+Q^{\dagger
}\right)\right]^2}
\left[\frac{{\partial }{\cal W}^{\dagger }}{\partial \left(\kappa
^{1/2}Q^{\dagger }\right)} +\frac{{\partial }{\cal W}}{\partial
\left(\kappa ^{1/2}Q\right)}\right]
+\sum _i\left\vert F_{z_i}\right\vert ^2\, .
\end{eqnarray}
where $F_{z_i}\equiv \left\langle {\rm e}^{\kappa K/2}
\left(\partial_{z_i} W + \kappa W\partial_{z_i}
K\right)\right\rangle$. The dynamics of the quintessence field is
determined by both the quintessence and hidden sectors. We also notice
that, as expected, the correction coming from the hidden sector is
proportional to the susy breaking mass $M_{_{\rm S}}$.

\subsection{The Soft Terms}
\label{subsec:soft}

Let us now turn to the calculation of the soft terms in the observable
sector. One usually obtains three types of terms. One is cubic in the
fields while the others are quadratic. In the present situation, this
property is clearly preserved. The new ingredient is that the soft
terms become quintessence dependent quantities. Following
Ref.~\cite{BMpart} and defining
\begin{eqnarray}
V_{_{\rm mSUGRA}} &=&\cdots + {\rm e}^{\kappa K}V_{\rm susy}+{\rm
e}^{\kappa K}A(Q)\lambda_{abc}\left(\phi_a \phi _b\phi
_c+\phi_a^{\dagger} \phi _a^{\dagger }\phi _c^{\dagger }\right) +{\rm
e}^{\kappa K}B(Q) \mu_{ab}\left(\phi _a \phi _b+\phi _a ^{\dagger
}\phi _b^{\dagger }\right) +m_{a\bar{b}}^2\phi _a \phi _{b}^{\dagger
}\, .
\end{eqnarray}
where the soft terms are the terms which are not in $V_{\rm susy}$,
one obtains for the $Q$--dependent coefficients $A$, $B$ and
$m_{a\bar{b}}$ in the noscale case
\begin{eqnarray}
\label{noscalesofta}
A(Q) &=& M_{_{\rm S}}\left(1+\frac13\sum _i\vert a_i\vert
^2+\frac13\sum _ia_ic_i\right)+\kappa M^3 \left[ {\cal W}^{\dagger
}\left(1+\frac13\sum _i\vert a_i\vert ^2\right)-\frac{1}{3} \kappa
^{1/2}\left(Q+Q^{\dagger }\right)\frac{{\partial }{\cal W}}{\partial
\left(\kappa ^{1/2}Q\right)}\right]\, , \\
\label{noscalesoftb}
B(Q) &=& M_{_{\rm
S}}\left(1+\frac12\sum _i\vert a_i\vert ^2+\frac12\sum
_ia_ic_i\right)+\kappa M^3 \left[ {\cal W}^{\dagger
}\left(1+\frac12\sum _i\vert a_i\vert ^2\right)-\frac{1}{2} \kappa
^{1/2}\left(Q+Q^{\dagger }\right)\frac{{\partial }{\cal W}}{\partial
\left(\kappa ^{1/2}Q\right)}\right]\, , \\
\label{noscalesoftm}
m_{a\bar{b}}^2(Q)
&=&\frac{{\rm e}^{\sum _i\vert a_i\vert ^2}}{ \left[\kappa
^{1/2}\left(Q+Q^{\dagger }\right)\right]^3}\left[M_{_{\rm S}}^2+\kappa
M_{_{\rm S}}M^3 \left({\cal W}+{\cal W}^{\dagger }\right)+\kappa
^2M^6{\cal W}{\cal W}^{\dagger }\right]\delta _{a\bar{b}}\, .
\end{eqnarray}
At this point, no assumption has been made except, of course, the
choice of the K\"ahler potential. However, it is clear that, in a
realistic model, we always have $M_{_{\rm S}}\gg \kappa M^3$ since the
susy breaking scale is much larger than the cosmological constant
scale, typically $M_{_{\rm S}}\sim 1 \mbox{TeV}$ while $\kappa M^6\sim
\left(10^{-3}\mbox{eV}\right)^4$. Now, the terms coming from $F_{z_i}$
in the scalar potential are of order $M_{\rm s}^2/\kappa$ which is
intolerably large compared to the cosmological scales. This is nothing
but another manifestation of the cosmological constant problem which,
again, is not solved in the framework of quintessence. This
contribution must be taken to vanish and therefore $a_i=c_i=0$.
Interestingly enough, it turns out to be exactly the case when $W_{\rm
hid}$ is a constant~\cite{BMcosmo}. Therefore $M_{_{\rm S}}$ is
constant, $A$ and $B$ are constant of the order of $M_{_{\rm S}}$, and
\begin{equation}
\label{uni}
2B=-M_{_{\rm S}}+3A\, ,
\end{equation}
while the mass $m_{a\bar{b}}$ acquires a very simple
$Q$-dependence given by
\begin{equation}
m_{a\bar{b}} =\frac{M_{_{\rm S}}}{\left[\kappa
^{1/2}\left(Q+Q^{\dagger }\right)\right]^{3/2}}\delta_{a\bar b}\,
.
\end{equation}
It is interesting to compare the above results to those obtained in
Ref.~\cite{BMcosmo} in the case of polynomial K\"ahler and super
potentials. The coefficients $A$ and $B$ were not constant but given
by $A=M_{_{\rm S}} \left(1+\kappa Q^2/3\right)$ and $B=M_{_{\rm S}}
\left(1+\kappa Q^2/2\right)$. We notice that, despite a different
dependence in the quintessence field, $A$ and $B$ also satisfy
Eq.~(\ref{uni}). On the other hand, the dependence of the soft term
$m_{a\bar{b}}$ is the same as in Ref.~\cite{BMcosmo}, namely
$m_{a\bar{b}}\propto M_{_{\rm S}}\exp\left(\kappa K/2\right)$. In the
SUGRA case this came from the fact that $\left \langle W_{\rm
quint}\right \rangle =0$ while in the no scale situation this
originates from neglecting subdominant terms thanks to the relation
$M_{_{\rm S}}\gg \kappa M^3$. However, since the K\"ahler potentials
are different, the above relation leads to different Q-dependence for
$m_{a\bar{b}}$.

\subsection{The Electro-Weak Transition in Presence of No-Scale
Quintessence}
\label{subsec:electroweak}

We now consider the application of the previous results to the
electroweak symmetry breaking since this is the way fermions in the
standard model are given a mass.  As is well-known, the potential in
the Higgs sector which belongs to the observable sector is modified by
the soft terms. Since these soft terms now depend on the quintessence
field, the Higgs potential also becomes a $Q$-dependent quantity. In
the MSSM, there are two $\mbox{SU}(2)_{\rm L}$ Higgs doublets
\begin{equation}
H_{\rm u}=\begin{pmatrix} H_{\rm u}^+ \cr H_{\rm u}^0
\end{pmatrix} \, , \quad H_{\rm d}=\begin{pmatrix} H_{\rm d}^0 \cr
H_{\rm u}^-
\end{pmatrix}\, ,
\end{equation}
that have opposite hypercharges, \ie $Y_{\rm u}=1$ and $Y_{\rm
d}=-1$. The only term which is relevant in the superpotential is
$W_{\rm obs}=\mu H_{\rm u}\cdot H_{\rm d}+\cdots$. This term gives
contribution to the globally susy term $V_{\rm susy}$ via the F-
and D-terms. Then, we have the contribution coming from the soft
susy-breaking terms. There is a B-soft susy-breaking term coming
from Eq.~(\ref{noscalesoftb}) and a contribution from the soft
masses, see Eq.~(\ref{noscalesoftm}). In order to evaluate the
latter, one writes $m_{1\bar{1}}=m_{H_{\rm u}}^2{\rm e}^{\kappa
K_{\rm quint}}$, and $m_{2\bar{2}}=m_{H_{\rm d}}^2{\rm e}^{\kappa
K_{\rm quint}}$, where $m_{H_{\rm u}}=m_{H_{\rm d}}=m_{3/2}^0$ at
the GUT scale. This degeneracy is lifted by the renormalisation
group evolution as necessary to obtain the radiative breaking of
the electroweak symmetry~\cite{Savoy}. The total Higgs potential,
taking $H_{\rm u}^0$ and $H_{\rm d}^0$ real since they have
opposite hypercharges, reads
\begin{eqnarray}
\label{simpleVhiggs} V^{_{\rm Higgs}}&=& {\rm e}^{\kappa K_{\rm
quint}}\biggl[\left(\left \vert \mu \right \vert ^2 +m_{H_{\rm
u}}^2\right)\left\vert H_{\rm u}^0\right\vert ^2+ \left(\left
\vert \mu \right \vert ^2 +m_{H_{\rm d}}^2\right)\left\vert H_{\rm
d}^0\right\vert ^2-2\mu B(Q)\left\vert H_{\rm u}^0\right\vert
\left\vert H_{\rm d}^0\right\vert\biggr]\nonumber \\ & & +\frac18
\left(g^2+g'^2\right) \left( \left\vert H_{\rm u}^0\right\vert
^2-\left\vert \ H_{\rm d}^0\right \vert^2\right )^2 \, .
\end{eqnarray}
The next step is to perform the minimization of the Higgs
potential given by Eq.~(\ref{simpleVhiggs}). In presence of dark
energy, the minimum becomes $Q$--dependent and the particles of
the standard model acquire a $Q$-dependent mass. Straightforward
calculations give
\begin{eqnarray}
\label{min1} {\rm e}^{\kappa K_{\rm quint}}\left(\left \vert \mu
\right \vert ^2+m_{H_{\rm
u}}^2\right) &=& \mu B(Q)\frac{{\rm e}^{\kappa K_{\rm
quint}}}{\tan \beta
}+\frac{m_{Z^0}^2}{2}\cos \left(2\beta \right)\, , \\
\label{min2} {\rm e}^{\kappa K_{\rm quint}}\left(\left \vert \mu
\right \vert ^2+m_{H_{\rm d}}^2\right)&=&\mu B(Q){\rm e}^{\kappa
K_{\rm quint}}\tan \beta -\frac{m_{Z^0}^2}{2}\cos \left(2\beta
\right)\, ,
\end{eqnarray}
where we have defined the Higgs vevs as $\langle H_{\rm u}^0\rangle
\equiv v_{\rm u}$, $\langle H_{\rm d}^0\rangle \equiv v_{\rm d}$,
$\tan \beta \equiv v_{\rm u}/v_{\rm d}$, or $v_{\rm u}=v \sin \beta $
and $v_{\rm d}=v\cos \beta $ and $m_{Z^0}$ as the gauge boson
$Z^0$. Adding the two equations for the minimum, we obtain a quadratic
equation determining $\tan \beta $. The solution can easily be found
and reads
\begin{eqnarray}
\label{tan}
\tan \beta (Q) &=& \frac{2\vert \mu \vert ^2+m_{H_{\rm
u}}^2(Q)+m_{H_{\rm d}}^2(Q)}{2\mu B(Q)}
\biggl(1 \pm \sqrt{1-4\mu ^2B^2(Q)\left[2\vert \mu \vert ^2+m_{H_{\rm
u}}^2(Q)+m_{H_{\rm d}}^2(Q)\right]^{-2}}\biggr)\, .
\end{eqnarray}
{\it A priori}, this equation is a transcendental equation
determining $\tan \beta $ as $\tan \beta $ also appears in the
right-hand-side of the above formula, more precisely in the Higgs
masses. Indeed, the two loop expression for the renormalized Higgs
masses gives~\cite{Brax}
\begin{eqnarray}
\label{mhu} m^2_{H_{\rm u}}\left( Q\right) &=& m^2_{H_{\rm d}}(Q)
-0.36\left(1+\frac{1}{\tan^2 \beta}\right)
\Biggl\{\left[m_{3/2}^0\left(Q\right)\right]^2\left(1-
\frac{1}{2\pi}\right)+8\left[m_{1/2}^0\left(Q\right)\right]^2
\nonumber \\ & & + \left(0.28 -\frac{0.72}{\tan ^2\beta }\right
)\left[A(Q) + 2 m^0_{1/2}\right]^2\Biggr\}\, ,\\
\label{mhd} m^2_{H_{\rm d}}\left(Q\right) &=&
      \left[m_{3/2}^0\left(Q\right)\right]^2\left(1-\frac{0.15}{4\pi}\right)
      + \frac{1}{2} \left[m^0_{1/2}\left(Q\right)\right]^2\, ,
\end{eqnarray}
where $m_{1/2}^0$ is the gaugino mass at GUT scale. However,
Eq.~(\ref{tan}) gives the leading order contribution of an expansion
in $1/\tan ^2 \beta $. As we have seen in the text, the noscale
situation is such that $A(Q)$ and $B(Q)$ are constant in $Q$ and,
therefore, the Higgs mass given by Eqs.~(\ref{mhu}), (\ref{mhd}) and
hence $\tan \beta $ do not depend on $Q$ in this particular
case. Again, this is very different from the polynomial case where
$\tan \beta $ is a $Q$-dependent quantity, see Eq.~(2.31) of
Ref.~\cite{BMcosmo} for the exact formula.

\par

From the equations~(\ref{min1}) and (\ref{min2}), one can also
deduce how the scale $v\equiv \sqrt{v_{\rm u}^2+v_{\rm d}^2}$
depends on the quintessence field. This leads to
\begin{eqnarray}
\label{generalv} v(Q)=\frac{2{\rm e}^{\kappa K_{\rm
quint}/2}}{\sqrt{g^2+g'{}^2}} \sqrt{\left\vert\left\vert \mu
\right \vert ^2+m_{H_{\rm u}}^2\right\vert}+{\cal
O}\left(\frac{1}{\tan
  \beta }\right)\, .
\end{eqnarray}
Again, the noscale case is quite particular: the only
$Q$--dependence is given by the factor $\exp\left(\kappa K_{\rm
quint}/2\right)$ in front of the whole expression.

\par

Then, finally, one has for the vevs of the two Higgs
\begin{eqnarray}
\label{vuvd} v_{\rm u}(Q)&=&\frac{v(Q)\tan \beta
(Q)}{\sqrt{1+\tan^2 \beta (Q)}}
=v(Q)+{\cal O}\left(\frac{1}{\tan ^2\beta }\right)\, ,\\
\label{vuvd2} v_{\rm d}(Q) &=& \frac{v(Q)}{\sqrt{1+\tan ^2\beta
(Q)}} =\frac{v(Q)}{\tan \beta }+{\cal O}\left(\frac{1}{\tan
^2\beta }\right)\, ,
\end{eqnarray}
at leading order in $1/\tan ^2\beta $ (but if we insert the
expression of $v$, then $v_{\rm u}$ and $v_{\rm d}$ are only
determined at first order in $1/\tan \beta $). This allows us to
deduce the two kinds of fermion masses, depending on whether the
fermions couple to $H_{\rm u}$ or $H_{\rm d}$
\begin{equation}
m_{{\rm u},a}^{_{\rm F}}(Q)= \lambda_{{\rm u},a}^{_{\rm F}} {\rm
  e}^{\kappa K_{\rm quint}/2}v_{\rm u}(Q)\,
  , \quad m_{{\rm d},a}^{_{\rm F}}(Q)=\lambda_{{\rm d},a}^{_{\rm F}}
  {\rm e}^{\kappa K_{\rm quint}/2}v_{\rm
  d}(Q)\, ,
\end{equation}
where $\lambda_{{\rm u},a}^{_{\rm F}}$ and $\lambda_{{\rm d},a}^{_{\rm
F}}$ are the Yukawa coupling of the particle $\phi_a$ coupling either
to $H_{\rm u}$ or $H_{\rm d}$. The masses pick up a $\exp\left(\kappa
K_{\rm quint}/2\right)$ dependence from the expression of $v(Q)$ and
another factor $\exp\left(\kappa K_{\rm quint}/2\right)$ from the
definition of the mass itself. As a result we have $m\propto
\exp\left(\kappa K_{\rm quint}\right) \propto Q^{-3}$ in the no scale
situation. This $Q$ dependence is the same for particles of type
``${\rm u}$'' or ``${\rm d}$'' as $\tan \beta $ is a constant. This
leads us to the main result of the section: in no scale quintessence
the behavior of the standard model particle masses is universal and
given by
\begin{equation}
\label{scalingmass}
m(Q)\propto \frac{1}{\left[\kappa ^{1/2}\left(Q+Q^{\dagger
}\right)\right]^3}\propto {\rm e}^{-\sqrt{6}q}\, .
\end{equation}
In the next section, we investigate the consequences of this
dependence for gravity experiments.

\section{Gravitational Tests and Chameleons}
\label{sec:testcham}

Let us now discuss the consequences of having $Q$-dependent
masses. This can lead to strong constraints coming from
gravitational experiments. Indeed, if the no-scale dark energy
potentials obtained in the previous sections, see
Eq.~(\ref{potDE}) for the quantity $V_{_{\rm DE}}$, are of the
runaway type (otherwise, in general, one can show that the
corresponding cosmological model is not interesting since it
becomes equivalent to the case of the cosmological constant, for a
specific example, see Ref.~\cite{BMcosmo}), then this implies that
the moduli have a mass $m_Q\sim H_0$, \ie of the order of the
Hubble rate now.  This implies that the range of the force
mediated by the quintessence field is large and, for instance, it
induces a fifth force and/or a violation of the weak equivalence
principle. In order to satisfy the constraints coming from fifth
force experiments such as the recent Cassini spacecraft
experiment, one must require that the Eddington (post-Newtonian)
parameter $\vert \gamma -1\vert \le 5\times 10^{-5}$, see
Ref.~\cite{GR}. If one defines the parameter $\alpha _{\rm u,d}$
by
\begin{equation}
\label{alpha} \alpha_{\rm u,d}(Q) \equiv \left\vert  \frac{{\rm
d}\ln m_{\rm u,d}^{_{\rm F}}(q)}{{\rm d} q} \right \vert \, ,
\end{equation}
where the derivative is taken with respect to the normalized field
$q$, then the difficulties are avoided by imposing that $\alpha_{\rm
u,d}^2\le 10^{-5}$ since one has $\gamma =1+2\alpha _{\rm u,d}^2$. In
our case, Eq.~(\ref{scalingmass}) implies
\begin{equation}
\alpha_{\rm u,d}= \sqrt{6}\, .
\end{equation}
This result is valid for a gedanken experiment involving the
gravitational effects on elementary particles. For macroscopic bodies,
the effects can be more subtle and will be discussed later, see also
Ref.~\cite{BMcosmo}. Of course, the above result is in contradiction
with the bounds on the existence of a fifth force and on the violation
of the weak equivalence principle.

\par

However, the above description is too naive because we have not taken
into account the chameleon effect in the presence of matter which, in
the framework used here, is necessarily present. Indeed, in the
presence of surrounding matter like the atmosphere or the
inter-planetary vacuum, the effective potential for the quintessence
field is modified by matter and becomes
\begin{equation}
\label{poteff} V_{\rm eff}(Q)= V_{_{\rm DE}}(Q) + A(Q) \rho_{\rm mat}\,
,
\end{equation}
where $A(Q)$ is the coupling of the quintessence field to matter, \ie
the mass of matter is proportional to $\propto A(Q)$. This can lead to
an effective minimum for the potential even though the Dark Energy
potential is runaway. In our case, see Eq.~(\ref{scalingmass}), we
have
\begin{equation}
V_{\rm eff}(Q)= V_{_{\rm DE}}(Q) + \left (\frac{Q_0}{Q} \right )^3
\rho_{\rm mat}=V_{_{\rm DE}}(q) + {\rm e}^{-\sqrt{6}(q_0-q)} \rho_{\rm
mat}\, ,
\end{equation}
where we have normalized the coupling to its present vacuum value
when $Q=Q_0$. For runaway potentials, the effective potential
possesses a minimum where
\begin{equation}
V'_{_{\rm DE}}\left(q_{\rm min}\right)= \sqrt 6 {\rm
e}^{-\sqrt{6}(q_0-q_{\rm min})} \rho_{\rm mat}\, ,
\end{equation}
and the mass at the minimum is
\begin{equation}
m^2_q= \kappa\left[V''_{_{\rm DE}} \left(q_{\rm min}\right) +\sqrt
6V'_{_{\rm DE}} \left(q_{\rm min}\right)\right]\, ,
\end{equation}
which is always of order $H_0$, \ie an almost massless field.  This
would lead an observable fifth force if it were not for the
possibility of a thin shell effect.

\par

Before turning to this question, it is worth commenting on the
chameleon effect in the SUGRA case, see Ref.~\cite{BMcosmo}. Since it
is a natural consequence of the couplings between the observable and
dark sector, the chameleon effect is also present in this
model. However, it is hidden by the susy breaking term $m_{3/2}^2Q^2$,
where $m_{3/2}$ is the gravitino mass which largely dominates the term
$A(Q)\rho _{\rm mat}$. In the no scale case, thanks to the very
particular form of the K\"ahler potential, the above susy breaking
term is not present and {\it a priori} the chameleon effect can be
efficient. In any case, in order to study whether no scale
quintessence is ruled out or not because of the gravity experiments,
it is mandatory to take into account the chameleon phenomenon
correctly.

\subsection{The Thin Shell Mechanism}
\label{subsec:cham}

A theory, as described before in this article, where the particle mass
depends on the quintessence field becomes a scalar tensor theory with
the Lagrangian
\begin{eqnarray}
\label{st} S &=&  \int {\rm d}^4 x \sqrt{-g}
\left[\frac{R}{2\kappa} -\frac12
  g^{\mu \nu}\partial_{\mu} q\partial _{\nu }q-V_{_{\rm
  DE}}(q)\right]+ S_{\rm mat}\left[\phi_{a}, A^2(q)
  g_{\mu\nu}\right] \, .
\end{eqnarray}
 Then, the geodesic equation can be written as
\begin{equation}
\label{geodesic}
\frac{{\rm d}^2x^{\mu }}{{\rm d}\tau^2}+\Gamma ^{\mu}_{\nu \lambda }
\frac{{\rm d}x^{\mu }}{{\rm d}\tau} \frac{{\rm d}x^{\mu }}{{\rm d}\tau
}+\alpha _q\frac{\partial q}{\partial x_{\mu }}=0\, ,
\end{equation}
where $\alpha _q\equiv \partial \ln A/\partial q$. In the above
equation, the Christoffel symbols are those associated with the
metric $g_{\mu \nu}$. The last term, which represents the new
force originating from the quintessence field, comes from the fact
that the geodesic equation is established for the metric appearing
in the matter Lagrangian. As is apparent from Eq.~(\ref{st}), this
one is given by $A^2(q)g_{\mu \nu}$ and the presence of the
$A^2(q)$ factor is responsible for the new term in
Eq.~(\ref{geodesic}). Analyzing this equation in the weak field
regime, one finds that the acceleration felt by a test particle is
given by
\begin{equation}
a=a_{_{\rm N}}-\alpha _q\frac{\partial q}{\partial r}\, ,
\end{equation}
where $a_{_{\rm N}}$ is the usual Newtonian acceleration (assuming a
spherical body, see below).

\par

\begin{figure*}
  \centering
  \includegraphics[width=0.6\textwidth,clip=true]{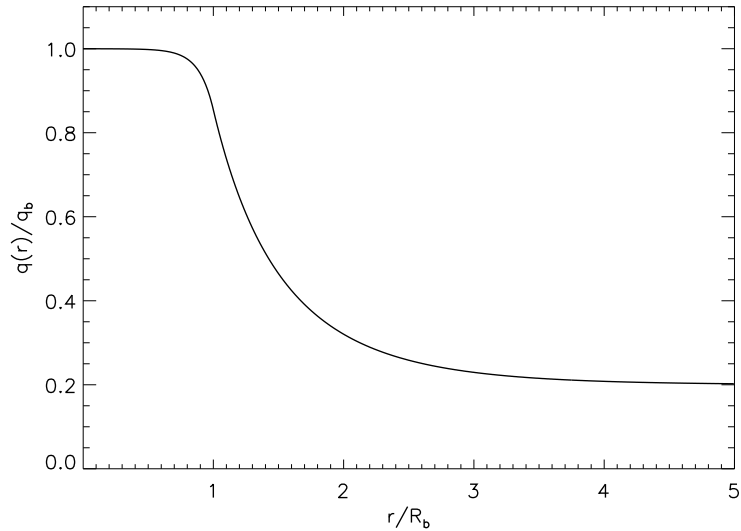}
  \caption[...]{Profile of the canonically normalized quintessence
field inside and outside a spherical body according to
Eqs.~(\ref{profileint}) and~(\ref{profileext}). As explained in the
text, $R_{\rm b}$ is the radius of the body and $q_{\rm b}$ is the
value of the quintessence field inside the body.}
\label{fig:profile}
\end{figure*}

Let us now consider a situation where the gravitational experiments
are performed on a body embedded in a surrounding medium. The body
could be a small ball of metal in the atmosphere or a planet in the
inter-planetary vacuum. The effective potential~(\ref{poteff}) is not
the same inside the body and outside because $\rho _{\rm matter}$ is
different. The effective potential can be approximated by
\begin{equation}
\label{approxVeff}
 V_{\rm eff}\simeq \frac12 m_q^2(q-q_{\rm min})^2\, ,
\end{equation}
where the minimum $q_{\rm min}$ is determined by $\partial V_{\rm
eff}/\partial q=0$ and the mass is $\partial ^2V_{\rm eff}/\partial
q^2$ evaluated at $q=q_{\rm min}$. As already mentioned the minimum
and the mass are different inside and outside the body. We denote by
$q_{\rm b}$ and $m_{\rm b}$ the minimum and the mass in the body and
by $q_{\infty}$ and $m_{\infty}$ the minimum and the mass of the
effective potential outside the body. Then, the Klein-Gordon equation
reads
\begin{equation}
\label{radialKG}
\frac{{\rm d}^2q}{{\rm d}r^2}+\frac{2}{r}\frac{{\rm d}q}{{\rm d}r}=
\frac{\partial V_{\rm eff}}{\partial q}\, ,
\end{equation}
where $r$ is a radial coordinate. Of course, the field $q$ should be
continuous at $r=R_{\rm b}$ where $R_{\rm b}$ is the radius of the
body. Notice that, in the Klein-Gordon equation, we have used
canonical kinetic terms in accordance with the fact that $q$ is a
canonically normalized field. With an effective potential given by
Eq.~(\ref{approxVeff}), the solution of Eq.~(\ref{radialKG}) reads
\begin{equation}
q=q_{\rm min}+\frac{A}{r}{\rm e}^{-mr}+\frac{B}{r}{\rm e}^{mr}\, ,
\end{equation}
where $A$ and $B$ are two arbitrary constant. Requiring that $q$
remains bounded inside and outside the body and joining the
interior and exterior solutions, one can determine the complete
profile which can be expressed as
\begin{eqnarray}
\label{profileint}
q_{<}\left(r\right) &=& q_{\rm b}+\frac{R_{\rm
b}\left(q_{\infty}-q_{\rm b}\right)\left(1+m_{\infty }R_{\rm
b}\right)}{\sinh \left(m_{\rm b}R_{\rm b}\right)\left[m_{\infty
}R_{\rm b}+m_{\rm b}R_{\rm b} \coth \left(m_{\rm b}R_{\rm
b}\right)\right]}\frac{\sinh \left(m_{\rm b}r\right)}{r}\, , \qquad
r\le R_{\rm b}\, , \\
\label{profileext}
q_{>}\left(r\right) &=& q_{\infty}+R_{\rm
b}\left(q_{\rm b}-q_{\infty}\right) \frac{m_{\rm b}R_{\rm
b}\coth\left(m_{\rm b}R_{\rm b}\right)-1}{\left[m_{\infty}R_{\rm
b}+m_{\rm b}R_{\rm b}\coth\left(m_{\rm b}R_{\rm
b}\right)\right]}\frac{{\rm e}^{-m_{\infty}\left(r-R_{\rm
b}\right)}}{r} \, , \qquad r\ge R_{\rm b}\, .
\end{eqnarray}
A typical profile is represented in Fig.~\ref{fig:profile}.

We are now in a position to estimate the acceleration caused by
the quintessence field. Assuming, as  is always the case in
practice, that $m_{\rm b}\gg m_{\infty}$, $m_{\rm b}R_{\rm b}\gg
1$, one has
\begin{equation}
\frac{\partial q_>(r)}{\partial r}\simeq -\frac{R_{\rm b}}{r^2}
\left(q_{\infty}-q_{\rm b}\right)\, ,
\end{equation}
from which we deduce that the acceleration felt by a test particle is
given by
\begin{equation}
a=-\frac{Gm_{\rm b}}{r^2}\left[1+\frac{\alpha
_q\left(q_{\infty}-q_{\rm b}\right)}{\Phi _{_{\rm N}}}\right]\, ,
\end{equation}
where $\Phi _{_{\rm N}}=Gm_{\rm b}/R_{\rm b}$ is the Newtonian
potential at the surface of the body. Therefore, the theory is
compatible with gravity tests if
\begin{equation}
\frac{\alpha _q\left(q_{\infty}-q_{\rm b}\right)}{\Phi _{_{\rm N}}}\ll
1\, .
\end{equation}
We see that the gravity tests are not sensitive to $\alpha _q$ but to
the combination $\alpha _q\left(q_{\infty}-q_{\rm b}\right)/\Phi
_{_{\rm N}}$. Hence, even if $\alpha _q$ is quite large, if the new
factor $\left(q_{\infty}-q_{\rm b}\right)/\Phi _{_{\rm N}}$ is small
then the model can be compatible. This is the thin shell effect.

\par

In our case, as $\alpha_q=\sqrt 6$, this implies that the moduli
fields must be small in order to satisfy the thin shell property. In
general, the Newton potential is very small, implying that the moduli
field $q$ must be small too. This strongly depends on the shape of the
potential and, therefore, on the superpotential in the moduli
sector. In the following we will give two examples which do not lead
to a thin shell. These examples have a well-motivated
superpotential. In non generic cases, no general obstruction to the
existence of a thin shell exists and, therefore, one may find moduli
superpotential leading both to quintessence and a thin-shell.

\subsection{Gaugino Condensation and Quintessence}
\label{subsec:kkltpot}

In order to go further, and to perform a quantitative calculation, one
must specify the dark energy potential which requires an explicit form
for the superpotential.

\par

In string compactifications, on top of the K\"ahler moduli there are
complex structure moduli and the string dilaton. These fields can be
stabilized once fluxes have been introduced. This leads to a
superpotential for the complex structure moduli and the dilaton. The
complex structure moduli and the dilaton lead to a supersymmetric
vacuum where they are fixed and the superpotential becomes a
constant. We are thus left with the K\"ahler moduli as a flat
direction. Once $D7$ branes are introduced in the setting,
non-perturbative gauge dynamics such as gaugino condensation implies
that a superpotential for the K\"ahler moduli is generated. On the
whole the dynamics of the K\"ahler moduli are governed by the
following superpotential~\cite{kklt}
\begin{equation}
W=M^3\left[w_0+c\exp\left(-\beta \kappa ^{1/2}Q\right)\right]\, ,
\end{equation}
where $w_0$, $c$ and $\beta $ are free and positive dimensionless
constants. It is immediate to find that the potential $V_{\rm quint}$
reads
\begin{equation}
V_{\rm quint}(Q)=\frac{\kappa M^6 c^2\beta }{2\left(\kappa
^{1/2}Q\right)^2} {\rm e}^{-\beta \kappa
^{1/2}Q}\left[\frac{w_0}{c}+{\rm e}^{-\beta\kappa
^{1/2}Q}\left(\frac{\beta }{3}\kappa ^{1/2}Q+1\right)\right]\, .
\end{equation}
Then, one should take into account the corrections coming from the
susy breaking terms. Using Eq.~(\ref{potDE}), one arrives at
\begin{equation}
V_{_{\rm DE}}(Q)=\frac{\kappa M^6 c^2\beta }{2\left(\kappa
^{1/2}Q\right)^2} {\rm e}^{-\beta \kappa
^{1/2}Q}\left[\frac{w_0}{c}+\frac{M_{_{\rm S}}}{c\kappa M^3}+{\rm
e}^{-\beta\kappa ^{1/2}Q}\left(\frac{\beta }{3}\kappa
^{1/2}Q+1\right)\right]\, .
\end{equation}
The effective potential has no minimum so no chameleon mechanism is
possible. Indeed, it is easy to demonstrate that $V_{_{\rm DE}}(Q)$ is
a decreasing function (for $\beta >0$ which is clearly the case of
physical interest) as $\exp\left(-\sqrt{6}q\right)$ is. Hence, this
model is ruled out gravitationally.

\subsection{Non-Renormalisable Potential}
\label{subsec:nrpot}

A class of potential with phenomenological interest can be
obtained if the quintessence field $Q$ has a non-renormalisable
superpotential. Although this is not what is expected from string
theory, we will consider as it leads to very appealing
quintessential properties. Therefore, we choose
\begin{equation}
W= -\frac{M^3}{n} (\kappa^{1/2} Q)^n\, ,
\end{equation}
Using Eq.~(\ref{potquint}), straightforward calculations lead to
the following form
\begin{equation}
V_{\rm quint}(Q)=\frac{\kappa M^6}{6n}(n-3)\left(\kappa
^{1/2}Q\right)^{2n-3} =\frac{\kappa M^6}{6n}(n-3)\exp
\left[-(2n-3)\sqrt{\frac23}q\right]\, .
\end{equation}
This leads to a satisfying exponential potential when $n>3$. These
potentials have been thoroughly studied and lead to the existence of
long time attractors~\cite{RP, quint}. Again, the runaway feature of
the potential implies that it is a good quintessence candidate. Then,
as expressed by Eq.~(\ref{potDE}), the shape of the quintessence
potential is modified by the soft terms present in the dark
sector. One obtains
\begin{eqnarray}
V_{_{\rm DE}}(Q)&=&\frac{\kappa M^6}{6n}(n-3)\left(\kappa
^{1/2}Q\right)^{2n-3} +\frac12M_{_{\rm S}}M^3 \left(\kappa
^{1/2}Q\right)^{n-3}\, , \\
&=& \frac{\kappa M^6}{6n}(n-3)\exp
\left[-(2n-3)\sqrt{\frac23}q\right] +\frac12M_{_{\rm S}}M^3\exp
\left[-(n-3)\sqrt{\frac23}q\right] \, .
\end{eqnarray}
As already discussed, the correction is proportional to the susy
breaking scale $M_{_{\rm S}}$. It has the structure of a two
exponential potential. As $q$ gets large, the second term of the
potential dominates and leads to acceleration in the matter era
provided $2/3 (n-3)^2< 4$ \ie $ n\le 3+\sqrt 6 $.  In this case, the
future of our Universe would be with $\Omega_Q=1$ with an equation of
state
\begin{equation}
w_Q= -1+ \frac{2(n-3)^2}{9}\, ,
\end{equation}
which is close to $-1$ when $n$ is close to $3$. Finally, the
effective potential for this model reads
\begin{equation}
V_{\rm eff}(q)= \frac12M_{_{\rm S}}M^3{\rm e}^{-(n-3)\sqrt{\frac23}q}
+ {\rm e}^{\sqrt{6}q} {\rm e}^{-\sqrt{6}q_0}\rho_{\rm mat}\, .
\end{equation}
From this expression one can deduce $q_{\rm min}$ and the mass of the
field at the minimum. One finds
\begin{equation}
q_{\rm min}=\frac{1}{n}\sqrt{\frac{3}{2}}\left\{\sqrt{6}q_0+\ln
\left[\frac{(n-3)M_{_{\rm S}}M^3}{6 \rho _{\rm mat}}\right]\right\}\,
, \quad m^2=\frac{n(n-3)}{3}\frac{M_{_{\rm S}}M^3}{\mP^2}
\left(\frac{n-3}{n}\frac{M_{_{\rm S}}M^3}{\rho _{\rm mat}}{\rm
e}^{\sqrt{6}q_0} \right)^{(3-n)/n}\, .
\end{equation}
As an example, let us consider the Earth in our local neighborhood
of the galaxy. We have $M_{_{\rm S}}\sim 10^{3}\mbox{GeV}$,
$M\simeq 10^{-12}\mbox{GeV}$, $\rho _{\oplus}\simeq 4\times
10^{-17}\mbox{GeV}^4$ and $\rho _{\infty} \simeq 4\times
10^{-42}\mbox{GeV}^4$. For $n=4$ and $q_0=1$, this gives
$q_{\oplus}\simeq -11.5$ and $q_{\infty}\simeq 6.1$. Since
$m_{\oplus}\simeq 3.3 \times 10^{51}\mbox{GeV}$ and
$R_{\oplus}\simeq 6\times 10^8 \mbox{cm}$, one gets $\Phi
_{\oplus}\simeq 1.4 \times 10^{-8}$ and therefore
\begin{equation}
\frac{\alpha _q\left(q_{\infty}-q_{\rm b}\right)}{\Phi _{_{\rm
N}}} \simeq 3\times 10^9 \gg1 \, .
\end{equation}
Since $m_{\infty}\simeq 2 \times 10^{-28}\mbox{eV}\ll
10^{-3}\mbox{eV}$, the range of the corresponding force is very
big. The conclusion is that, although we have a chameleon
mechanism which renders the analysis of the gravity tests non
trivial, this one is not efficient enough and the corresponding
model  is ruled out.

\section{Conclusion: a No-Go Theorem for Quintessence?}

We have presented models of moduli quintessence. Despite the large
gravitational coupling of the moduli to matter in these models, a
chameleon mechanism is at play and could render the models compatible
with gravitational experiments. Unfortunately, in realistic cases such
as gaugino condensation or non-renormalisable superpotentials, the
chameleon phenomenon is not strong enough to save the models.

\par

One can deduce a no-go theorem (modulo, of course, the assumptions
made in this article, in particular that of the separate sectors)
showing the incompatibility between quintessence in supergravity and
gravity tests. Let us come back to the general structure of the scalar
potential. As shown in Ref.~\cite{BMpart}, see Eq.~(2.18), it can
always be written as
\begin{equation}
V_{_{\rm DE}}(Q)= \kappa M^6v_1(\kappa ^{1/2}Q) +M_{_{\rm
S}}M^3v_2(\kappa ^{1/2}Q) +\frac{M_{_{\rm S}}^2}{\kappa } {\rm
e}^{\kappa K}\left(\kappa K^{QQ^{\dagger }} K_Q K_{Q^{\dagger }}
-3\right) + \sum_i \left\vert F_{z_i}\right \vert^2\, ,
\end{equation}
where we have chosen to emphasize the various combinations of scales
appearing in this expression and where, consequently, $v_1(\kappa
^{1/2}Q)$ and $v_2(\kappa ^{1/2}Q)$ are dimensionless functions, {\it
a priori} of order one at present time. The last term contains the
F-terms of the hidden sector.

\par

Let us consider first models where the K\"ahler potential can be
expanded around $Q=0$. After a K\"ahler transformation, one can always
expand
\begin{equation}
K_{\rm quint}= QQ^{\dagger } + \cdots \, ,
\end{equation}
where $\cdots$ represent Planck suppressed operators which, at present
time, are not necessarily negligible since we have $\left\langle
Q\right \rangle \sim \mP$ now. It is immediate to see that at leading
order, the quintessence field picks up a soft breaking mass
\begin{equation}
\label{depoly}
V_{_{\rm DE}}(Q)=\kappa M^6v_1(\kappa ^{1/2}Q) +M_{_{\rm
S}}M^3v_2(\kappa ^{1/2}Q) + m_{3/2}^2 \vert Q\vert^2\sim M_{_{\rm
S}}M^3v_2(\kappa ^{1/2}Q) + m_{3/2}^2 \vert Q\vert^2\, ,
\end{equation}
where we have used that $M_{_{\rm S}}\propto m_{3/2}$ and have imposed
$\sum_i \vert F_{z_i}\vert^2= 3m_{3/2}^2\kappa^{-1}$ in order to
cancel the intolerably large contribution to the cosmological constant
coming from the hidden sector. The last equality originates from the
condition $M_{_{\rm S}}\gg \kappa M^3$. From Eq.~(\ref{depoly}), we
see that the potential acquires a minimum since, in general, the
functions $v_1$ and $v_2$ are of the runaway type, \ie decreasing with
$Q$. The value of the minimum is controlled by the scales $M$,
$M_{_{\rm S}}$ and $m_{3/2}$. Due to the large value of $m_{3/2}$
compared to the quintessence field, the minimum is generically small
in Planck units. The scale $M$ is tuned to get a minimum value for the
potential of order $\Omega_{_{\Lambda }}\rho_{\rm cri}$. At this
minimum, the mass of the quintessence field is $m_{3/2}$ large enough
to evade all the gravitational tests. Now, cosmologically, the
steepness of the quadratic potential in $Q$ implies that the field
must have settled at the minimum before Big Bang Nucleosynthesis
(BBN). If not, the energy density of the quintessence field would
exceed the MeV energy scale of BBN. In practice, the potential is
constant since BBN, \ie equivalent to a cosmological constant. Notice
that the coupling of the quintessence field to matter induces a
correction to the potential in $\kappa \rho_{\rm mat} \vert Q\vert^2$
which is negligible compared to $m_{3/2} \vert Q\vert ^2$, hence no
chameleon effect.

\par

One can circumvent this argument by taking singular potentials
where the potential term in $\vert W\vert^2$ is constant. One can
choose
\begin{equation}
K_{\rm quint}=-\frac{n}{\kappa} \ln
\left[\kappa^{1/2}\left(Q+Q^{\dagger }\right)\right]\, .
\end{equation}
In this case, n=3 for moduli and n=1 for the dilaton. Fine-tuning
of the cosmological constant requires
\begin{equation}
\sum_i \vert F_{z_i}\vert^2= (3-n) m_{3/2}^2\kappa^{-1}\, ,
\end{equation}
leaving
\begin{equation}
V_{_{\rm DE}}(Q)=\kappa M^6v_1(\kappa ^{1/2}Q) +M_{_{\rm
S}}M^3v_2(\kappa ^{1/2}Q)\sim M_{_{\rm S}}M^3v_2(\kappa ^{1/2}Q)\, .
\end{equation}
No mass term appears for the quintessence field. The dynamics are
similar to the no-scale case with a contribution from the matter
density. The mass of the quintessence field at the minimum of the
matter-dependent potential is of order $H_0$. Moreover the thin-shell
effect is only present for small values of the normalized scalar field
$q$, a situation which requires a non-generic quintessence
superpotential (otherwise $q\sim 1$ generically).

\par

We conclude that under broad circumstances, one cannot obtain a
compatibility between quintessence and gravity tests in
supergravity. Either the dynamics are equivalent to a cosmological
constant or gravity tests are not evaded. One possibility is to
relinquish the assumption on the K\"ahler potential (three decoupled
K\"ahler potentials). Work on this possibility (sequestered models and
others) is in progress.

\label{sec:conclusion}



\end{document}